\documentclass[aps,eqsecnum,tightenlines]{revtex4}
\usepackage{graphicx}
\usepackage{bm}

\begin{document}
\hskip5.5in{NT@UW-02-001}

\title{Return of the EMC effect: finite nuclei}
\author{Jason R. Smith}
\author{Gerald A. Miller}
\affiliation{Department of Physics\\
University of Washington\\Seattle, WA 98195-1560}
\begin{abstract}
A light front formalism for deep inelastic lepton scattering from
finite nuclei is developed. In particular, the nucleon plus
momentum distribution and a finite system analog of the
Hugenholtz-van Hove theorem are presented. Using a relativistic
mean field model, numerical results for the plus momentum
distribution and ratio of bound to free nucleon structure
functions for Oxygen, Calcium and Lead are given. We show that we
can incorporate light front physics with excellent accuracy while
using easily computed equal time wavefunctions. Assuming nucleon
structure is not modified in-medium we find that the calculations
are not consistent with the binding effect apparent in the data
not only in the magnitude of the effect, but in the dependence on
the number of nucleons.
\end{abstract}
\maketitle

\section{Introduction}
\label{sec:intro}

The  nuclear structure function $F_{2A}(x)$ is smaller than $A$
times the free nucleon structure function $AF_{2N}(x)$ for values
of $x$ in the regime where valence quarks are dominant. This
phenomenon, known as the European Muon Collaboration (EMC) effect
\cite{Aubert:1983xm}, has been known for almost twenty years.
Nevertheless, the significance of this observation remains
unresolved even though there is a clear interpretation within the
parton model: a valence quark in a bound nucleon carries less
momentum than a  valence quark in a free one. There are many
possible explanations, but no universally accepted one.  The
underlying  mechanism responsible for the transfer of momentum
within the constituents of the nucleus has not yet been specified.
One popular mechanism involves ordinary nuclear binding which, in
its simplest inculcation, is represented by evaluating the free
nucleon structure function at a value of $x$ increased by a factor
of the average separation energy divided by the nucleon mass $
\bar{\epsilon}/M \simeq 0.04$. The validity of this binding effect
has been questioned; see the reviews
\cite{Arneodo:1992wf,Geesaman:1995yd,Piller:1999wx,Frankfurt:nt}.

The Bjorken variable $x$ is a fraction of the  plus component of
momentum, and the desire to obtain a more precise evaluation and
understanding of the binding effect lead us to attempt to obtain a
nuclear wave function in which the momentum of the nucleons is
expressed in terms of this same plus component. Therefore we
applied light front dynamics to determining nuclear wave functions
\cite{previous}. In this formalism one defines $x^{\pm}=x^{0}\pm
x^{3}$ and quantizes on equal $x^{+}$ surfaces which have a
constant light front time, $\tau$. The   conjugate operator
$P^{-}$ acts as an evolution operator for $\tau$. The plus
momentum is canonically conjugate to the spatial $x^{-}$ variable.
This light front formalism has a variety of advantages
\cite{lcrev1,lcrev2,lcrev3,lcrev4,lcrev5,lcrev6,lcrev7,lcrev8,lcrev9}
and also entails complications \cite{complications}.

Our most recent result \cite{Miller:2001tg} is that the use of the
relativistic mean field approximation, and the assumption that the
structure of the nucleon is not modified by effects of the medium,
to describe infinite nuclear matter leads to no appreciable
binding effect. The failure was encapsulated in terms of the
Hugenholtz-van Hove theorem \cite{HvH} which states that the
average nuclear binding energy per nucleon is equal to the binding
energy of a nucleon at the top of the Fermi sea. The light front
version of this theorem is obtained from the requirement that, in
the nuclear rest frame, the expectation values of the total plus
and minus momentum are equal. The original version of the theorem
was obtained in a non-relativistic theory in which nucleons are
the only degrees of freedom. Here, the mesons are important and
the theory is relativistic, but the theorem still holds. This
theorem can be shown to restrict \cite{Miller:2001tg} the plus
momentum carried by nucleons to be the mass of the nucleus, which
in turn implies that the probability for a nucleon to have a plus
momentum $k^{+}$ is narrowly peaked about
$k^{+}=M_{A}/A=\overline{M}$. Thus the only binding effect arises
from the average binding energy, which is much smaller than the
average separation energy. Therefore dynamics beyond the
relativistic mean field approximation must be invoked to explain
the EMC effect. This conclusion was limited to the case of
infinite nuclear matter, and the computed nuclear structure
function could only be compared with data on finite nuclear
targets extrapolated to the limit $A\to\infty$. The goal of the
present work is to extend the results to finite nuclei; the main
complication arises from the spatial dependence of the nucleon and
meson fields.

We briefly outline our procedure. In Sections \ref{sec:green} and
\ref{sec:subtle} we present the covariant deep inelastic
scattering formalism of Ref.~\cite{Jung:1988jw} and derive its
representation in terms of nucleon single particle wave functions.
The plus momentum distribution follows from this representation in
Section \ref{sec:derivation} where we also derive new version of
the Hugenholtz-van Hove theorem. Then we present the results of
analytic and numerical calculations in Section \ref{sec:numbers},
the latter giving an $A$ dependence of the ratio function contrary
to experimental results. This again gives the result that the use
of the relativistic mean field approximation, combined with the
assumption that the nuclear medium does not modify the structure
of the nucleon, cannot describe the EMC effect. The reasons for
the subtle differences between the results for finite nuclei and
nuclear matter are detailed in Section \ref{sec:scalar}. Finally,
we summarize and discuss possible implications.

\section{Nucleon Green's Function for Finite Nuclei}
\label{sec:green}

\begin{figure}
\centering
\includegraphics[scale=1.0]{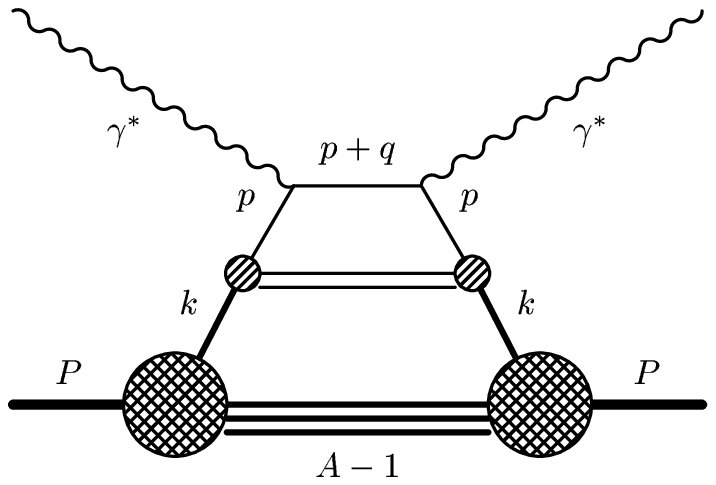}
\caption{Feynman diagram for deep inelastic scattering. A nucleus
of momentum $P$ is struck by a virtual photon of momentum $q$. We
label nucleon momentum $k$, and quark momentum $p$.}
\label{fig:dis}
\end{figure}

We begin with the covariant plus momentum distribution function
\begin{equation}
f_N(y)=\int {d^4k\over (2\pi)^4} \delta\left(y-\frac{k^0+k^3}
{\overline{M}}\right) \text{Tr} \left[ \frac{\gamma^{+}}{2P^{+}A}
\chi^{A}(k,P)\right] \label{eq:fn}
\end{equation}
where we identify
\begin{eqnarray}
\chi^{A}(k,P) & \equiv & -i \int d^{4}x \int d^{4}y e^{-ik\cdot
(x-y)} G^C(x,y),\label{eq:chi}
\end{eqnarray}
where $G^{C}(x,y)$ is the connected part of the nucleon Green's
function:
\begin{eqnarray}
iG(x,y) & \equiv &  \langle P | T^{+} \{\psi'(x) \overline{\psi}'
(y)\} |P \rangle. \label{eq:green}
\end{eqnarray}
This result is directly determined from the Feynman diagram in
Fig.~\ref{fig:dis}
following Ref.~\cite{Jung:1988jw}. So far this is independent of
the particular relativistic mean field model, but for concreteness
we use a Quantum Hadrodynamics (QHD) Lagrangian
\cite{Serot:1997xg,Serot:1986ey}, specifically QHD-I as in
Ref.~\cite{Blunden:1999gq}, where the nucleon fields, $\psi'$,
that appear in Eq.~(\ref{eq:green}) are those appearing in the
Lagrangian. Light front quantization requires that the plus
component of all vector potential fields vanishes, and this is
obtained by using the Soper-Yan transformation
\cite{Soper:1971wn,Yan:1973qf}
\begin{equation}
\psi'(x)\equiv e^{-ig_v\Lambda(x)}\psi(x),\qquad
\partial^{+}\Lambda(x)=V^{+}(x). \label{eq:soperyan}
\end{equation}
to define the nucleon field operator $\psi$ for various models
\cite{Miller:2001tg}. This transformation allows the use of the
eigenmode expansion for the $\psi$ fields which have been obtained
previously in Ref.~\cite{Blunden:1999gq}
\begin{eqnarray}
\psi(x) & = & \sum_{\alpha}  \left[ a_{\alpha} u_{\alpha} (x) +
b_{\alpha}^{\dag} v_{\alpha} (x) \right]\nonumber \\
& = & \sum_{\alpha}  \left[ a_{\alpha} u_{\alpha} (\bm{x}) e^{-i
p^{-}_{\alpha} x^{+}/2 }+ b_{\alpha}^{\dag} v_{\alpha} (\bm{x})
e^{i p^{-}_{\alpha} x^{+}/2 } \right], \label{eq:mode-exp}
\end{eqnarray}
where $a_{\alpha}$ and $b^{\dag}_{\alpha}$ are (anti-)nucleon
annihilation operators and we define $z \equiv -x^{-}/2$ with
$\partial^{+} = 2\partial_{-} = -\partial_{z} $ and
$(\bm{x}_{\bot},z) \equiv \bm{x}$ which allows us to treat the
minus and perpendicular co\"{o}rdinates on equal footing. The
$u_{\alpha}$ and $v_{\alpha}$ are co\"{o}rdinate space 4-component
spinor solutions to the light front Dirac equation with
eigenvalues $p^{-}_{\alpha}/2 = M-\varepsilon_{\alpha}$. To
simplify the analysis we will temporarily ignore electromagnetic
effects, but we will include them in the final numerical results.
The light front mode equations in QHD-I are obtained by minimizing
the $P^{-}$ operator (light front Hamiltonian) with the constraint
\cite{Blunden:1999gq} that $P^{+}=P^{-}$. The result is
\begin{eqnarray}
-i\partial_{z}|\psi_{\alpha}^{-}\rangle & = &
\left[\bm{\alpha}_{\bot}\cdot(\bm{p}_{\bot}-g_{v}\bar{\bm{V}}_{\bot})+\beta(M+g_{s}\phi)\right]
|\psi_{\alpha}^{+}\rangle \label{eq:dirac1} \\
p^{-}_{\alpha}|\psi_{\alpha}^{+}\rangle & = &
\left[-i\partial_{z}+2g_{v}\bar{V}^{-}\right]|\psi_{\alpha}^{+}\rangle
\nonumber\\* & & +
\left[\bm{\alpha}_{\bot}\cdot(\bm{p}_{\bot}-g_{v}\bar{\bm{V}}_{\bot})+\beta(M+g_{s}\phi)\right]
|\psi_{\alpha}^{-}\rangle, \label{eq:dirac2}
\end{eqnarray}
with
\begin{eqnarray}
\quad u_{\alpha}(\bm{x}) & = & \sqrt{2P^{+}A} \; \langle
\bm{x}|\psi_{\alpha}\rangle\\* \Lambda_{\pm}|\psi_{\alpha}\rangle
& = & \frac{1}{2}\gamma^{0}\gamma^{\pm}|\psi_{\alpha}\rangle =
|\psi_{\alpha}^{\pm}\rangle\\*
\partial^{+}\bar{V}^{\mu} & = & \partial^{+}V^{\mu} -
\partial^{\mu}V^{+}.
\end{eqnarray}
Using standard manipulations \cite{Serot:1986ey} and defining
$\varepsilon_{F}$ as the energy of the highest occupied state, we
find the Green's function to be
\begin{widetext}
\begin{eqnarray}
G(x,y) & = & \sum_{\alpha} u_{\alpha} (\bm{x})
\overline{u}_{\alpha}
(\bm{y})e^{-ig_{v}[\Lambda(\bm{x})-\Lambda(\bm{y})]} \int
\frac{dk^{-}}{2\pi} e ^{-ik^{-}(x^{+}-y^{+})/2} \left[
\frac{1}{k^{-}-p^{-}_{\alpha} + i\varepsilon} + 2\pi i
\delta(k^{-}-p^{-}_{\alpha})
\theta(\varepsilon_{F}-\varepsilon_{\alpha}) \right]\nonumber\\
& \equiv & G^{D}(x,y) + G^{C}(x,y),
\end{eqnarray}
\end{widetext}
where the superscripts $D$ and $C$ represent the disconnected and
connected parts of the nucleon Green's function, respectively. The
connected part is relevant to deep inelastic scattering and is
given by
\begin{equation}
G^{C}(x,y) = i \sum_{\alpha\in F} u_{\alpha} (\bm{x})
\overline{u}_{\alpha} (\bm{y})
e^{-ig_{v}[\Lambda(\bm{x})-\Lambda(\bm{y})]}
e^{-ip^{-}_{\alpha}(x^{+}-y^{+})/2}, \label{eq:cgreen}
\end{equation}
where the sum is over occupied levels $\alpha$ in the Fermi sea
$F$. We now substitute Eq.~(\ref{eq:cgreen}) into
Eq.~(\ref{eq:chi}), first defining $(\bm{k}_{\bot},k^{+}) \equiv
\bm{k}$ where $\bm{k}\cdot \bm{x} = \bm{k}_{\bot}\cdot
\bm{x}_{\bot} + k^{+}z  = \bm{k}_{\bot}\cdot \bm{x}_{\bot} -
k^{+}x^{-}/2 $, $d\bm{x}=d^{2}\bm{x}_{\bot}dz$,
$d\bm{k}=d^{2}\bm{k}_{\bot}dk^{+}$ and
\begin{equation}
u''_{\alpha}(\bm{k}) \equiv \int d\bm{x}
e^{-i\bm{k}\cdot\bm{x}}e^{-ig_{v}\Lambda(\bm{x})}u_{\alpha}(\bm{x}).
\label{eq:uprime}
\end{equation}
We find
\begin{equation}
\chi^{A}(k,P) = (2\pi)^{2}\sum_{\alpha\in F}
u''_{\alpha}(\bm{k})\overline{u}''_{\alpha}(\bm{k})
\delta(k^{-}-p^{-}_{\alpha}). \label{eq:new-chi}
\end{equation}
The motivation for the `double-prime' notation is the subject of
the next section.

\section{Wave function Subtleties}
\label{sec:subtle}

It would be useful to express $\chi^{A}(k,P)$ in terms of
solutions of the ordinary Dirac equation, because one may use a
standard computer program \cite{cjht}. To this end it is
convenient rewrite Eq.~(\ref{eq:uprime})
\begin{equation}
\langle \bm{k} |\psi''^{+}_{\alpha}\rangle = \int d\bm{x}
e^{-i\bm{k}\cdot\bm{x}}e^{-ig_{v}\Lambda(\bm{x})} \langle \bm{x}
|\psi^{+}_{\alpha}\rangle. \label{eq:tranpsi}
\end{equation}
Note that the difference between $|\psi''^{+}_{\alpha}\rangle $ of
Eq.~(\ref{eq:tranpsi}) and $u''_{\alpha}$ of Eq.~(\ref{eq:uprime})
is simply the normalization factor $\sqrt{2P^{+}A}$.
\begin{widetext}
These `double-primed' fields satisfy another version of the mode
equations Eq.~(\ref{eq:dirac1}) and Eq.~(\ref{eq:dirac2})
following from an application of the Soper-Yan transformation
Eq.~(\ref{eq:soperyan}), and are given by
\begin{eqnarray}
\left[-i\partial_{z}-g_{v}V^{+}\right]|\psi_{\alpha}''^{-}\rangle
& = & \left[\bm{\alpha}_{\bot}\cdot(\bm{p}_{\bot}
-g_{v}\bm{V}_{\bot}) + \beta(M+g_{s}\phi)\right]
|\psi_{\alpha}''^{+}\rangle \label{eq:dirac1-primed} \\*
\left[i\partial_{z}+g_{v}V^{+}-2g_{v}\bar{V}^{-}+p^{-}_{\alpha}\right]|\psi_{\alpha}''^{+}\rangle
& = & \left[\bm{\alpha}_{\bot}\cdot(\bm{p}_{\bot}
-g_{v}\bm{V}_{\bot}) + \beta(M+g_{s}\phi)\right]
|\psi_{\alpha}''^{-}\rangle. \label{eq:dirac2-primed}
\end{eqnarray}
If one multiplies Eq.~(\ref{eq:dirac1-primed}) by $\gamma^{+}$ and
Eq.~(\ref{eq:dirac2-primed}) by $\gamma^{-}$ and adds the two
equations, using $V^{+}=V^{-}=\bar{V}^{-}=V^{0}$, one obtains
\begin{equation}
\left[-\gamma^{3}(i\partial_{z}+p^{-}_{\alpha}/2)+\gamma^{0}
(p^{-}_{\alpha}/2-g_{v}V^{0}) \right] \psi_{\alpha}''(\bm{x}) =
\left[ \bm{\gamma}_{\bot}\cdot\bm{p}_{\bot} + M+g_{s}\phi
\right]\psi_{\alpha}''(\bm{x}) \label{eq:almost-dirac}
\end{equation}
Eq.~(\ref{eq:almost-dirac}) is almost the same as the Dirac
equation of the equal time formulation (for the $\psi'$ fields),
with the exception of the $p^{-}_{\alpha}$ term multiplying
$\gamma^{3}$. Removing the offending term gives the relationship
between the $\psi'$ and $\psi''$ fields
\begin{eqnarray}
\langle \bm{x}| \psi'_{\alpha} \rangle & \simeq &
e^{-ip^{-}_{\alpha}z/2}\langle \bm{x}| \psi''_{\alpha} \rangle
\label{eq:prime-doubleprime}\\* & = & e^{-ip^{-}_{\alpha}z/2}
e^{-ig_{v}\Lambda(\bm{x})}\langle \bm{x}| \psi_{\alpha} \rangle
\label{eq:prime-noprime}
\end{eqnarray}
Eq.~(\ref{eq:prime-noprime}) is the approximate relationship
between the $\psi$ and $\psi'$ fields determined in
Ref.~\cite{Blunden:1999gq}. The approximation lies in the fact
that the spectrum condition is not maintained exactly, and the
resulting Fourier transform of the wavefunction will have
unphysical support for $k^{+}<0$. This support is largely
irrelevant as it manifests far out on an exponential tail since
$p^{-}_{\alpha}$ contains the nucleon mass. The relationship
between the $\psi$ and $\psi''$ is exactly our definition
Eq.~(\ref{eq:tranpsi}). The use of
Eq.~(\ref{eq:prime-doubleprime}) in Eq.~(\ref{eq:almost-dirac})
leads immediately to the result that the fields  $\psi'_{\alpha}$
satisfy the ordinary Dirac equation
\begin{equation}
\gamma^{0} (p^{-}_{\alpha}/2-g_{v}V^{0}) \psi_{\alpha}'(\bm{x}) =
\left[ \bm{\gamma}\cdot\bm{p} + M+g_{s}\phi
\right]\psi_{\alpha}'(\bm{x})
\end{equation}
\end{widetext}
We now are ready to derive a representation of Eq.~(\ref{eq:fn})
in terms of these nucleon wave functions.

\section{Derivation of the Plus Momentum Distribution}
\label{sec:derivation}

In Ref.~\cite{Blunden:1999gq}, it was determined that a plus
momentum distribution in QHD-I is given by
\begin{equation}
f(k^{+}) = 2 \sum_{\alpha\in F} \int d^{2}\bm{x}_{\bot}
\left|\langle \bm{x}_{\bot},k^{+}|\psi^{+}_{\alpha}\rangle \right|
^{2}. \label{eq:flf}
\end{equation}
This distribution peaks at $k^{+}/\overline{M}\equiv y \simeq 0.8$
for $^{16}$O, (with smaller values for heavier nuclei) but is  not
the distribution obtained from the covariant formalism of Section
\ref{sec:green}. The connection between this $f(y)$ and the
covariant $f_{N}(y)$ was made in Ref.~\cite{Miller:2001tg}; it was
determined that, in the limit of infinite nuclear matter,  the
relationship between $f(y)$ and $f_{N}(y)$ is  simply a shift in
the argument by the vector meson potential:
\begin{equation}
f(y) = f_{N}(y+g_{v}V^{+}/\overline{M}). \label{eq:shift-nm}
\end{equation}
This shift arises from the use of the Soper-Yan transformation
Eq.~(\ref{eq:soperyan}) where the $\psi'$ fields are those
appearing in the Lagrangian and are used to determine $f_{N}(y)$,
whereas the $\psi$ fields are used to determine $f(y)$. In finite
nuclei, this relationship is somewhat more complicated since the
vector meson potential is no longer a constant over all space. We
start with Eq.~(\ref{eq:new-chi}), and see that
\begin{eqnarray}
\text{Tr}\gamma^{+}\chi^{A}(k,P) & = &  (2\pi)^{2} \sum_{\alpha\in
F} \text{Tr} \left[\gamma^{+}u''_{\alpha}(\bm{k})
\overline{u}''_{\alpha}(\bm{k})\right]
\delta(k^{-}-p^{-}_{\alpha})\nonumber\\* & = & 16\pi^{2} P ^{+} A
\sum_{\alpha\in F} \left|\langle \bm{k}|\psi''^{+}_{\alpha}\rangle
\right|^{2} \delta(k^{-}-p^{-}_{\alpha}).\nonumber
\end{eqnarray}
Substituting into Eq.~(\ref{eq:fn}) we obtain
\begin{eqnarray}
f_N(y) & = & \frac{2}{(2\pi)^{2}} \sum_{\alpha\in F} \int d\bm{k}
\delta(y-k^{+}/\overline{M}) \left|\langle \bm{k}
|\psi''^{+}_{\alpha}\rangle \right|^{2}. \label{eq:fny}
\end{eqnarray}
Use of Parseval's identity and integrating over $k^{+}$ gives us
our main result:
\begin{eqnarray}
f_N(y) & = & 2\overline{M} \sum_{\alpha\in F} \int
d^{2}\bm{x}_{\bot} \left|\langle \bm{x}_{\bot}, \overline{M}y
|\psi''^{+}_{\alpha}\rangle \right|^{2}, \label{eq:fn-general}
\end{eqnarray}
so the plus momentum distribution is related to Fourier transform
of the $\psi''$ wave functions. One can see the similarity to
Eq.~(\ref{eq:flf}); the difference lies entirely in
Eq.~(\ref{eq:tranpsi}). It should be emphasized that this result
does not depend on the approximation in Section \ref{sec:subtle}.

We shall use $f_{N}(y)$ to compute the nuclear structure function
$F_{2A}(x)$ in Section \ref{sec:numbers}, but first we derive a
version of the Hugenholtz-van Hove theorem valid for finite
nuclei. To do that, multiply Eq.~(\ref{eq:fny}) by $y$ and
integrate
\begin{eqnarray}
\langle y \rangle & \equiv & \int dy y f_{N}(y)\nonumber\\* & = &
\frac{2}{(2\pi)^{2}} \sum_{\alpha\in F}\int d\bm{k}
\frac{k^{+}}{\overline{M}} \left|\langle \bm{k}
|\psi''^{+}_{\alpha}\rangle \right|^{2},
\end{eqnarray}
Now remove the plus projections and re-express $\psi''$ and its
complex conjugate in co\"{o}rdinate spaces $\bm{x}$ and $\bm{x'}$.
One can then integrate over $\bm{k}$ yielding a delta function
$\delta (\bm{x}-\bm{x'})$ which allows integration over $\bm{x'}$
\begin{eqnarray}
\langle y \rangle & = & \frac{1}{\overline{M}} \sum_{\alpha\in
F}\int d\bm{x} \psi''^{\dag}_{\alpha}(\bm{x}) \gamma^{0}
\gamma^{+} i\partial^{+} \psi_{\alpha}''(\bm{x}) \nonumber
\end{eqnarray}
We wish to look at the $\psi$ fields in order to understand our
result in the context of Ref.~\cite{Blunden:1999gq}, so we need to
perform the Soper-Yan transformation Eq.~(\ref{eq:soperyan}) and
use $X^{\dag}\gamma^{0}\equiv\overline{X}$
\begin{eqnarray}
\langle y \rangle & = & \frac{1}{\overline{M}} \sum_{\alpha\in
F}\int d\bm{x} \overline{\psi}_{\alpha}(\bm{x}) \gamma^{+}
[i\partial^{+}+g_{v}V^{+}(\bm{x})] \psi_{\alpha}(\bm{x}) \nonumber
\end{eqnarray}
If we explicitly put in the the nuclear state vectors, we can
perform the sum on $\alpha$ by inserting creation and annihilation
operators; we can add the time dependence for free since it is
unaffected by $\partial^{+}$ and cancels with both fermion fields,
and the vector potential is static. We have effectively undone the
substitution Eq.~(\ref{eq:mode-exp}) and now have an expectation
value of an operator
\begin{eqnarray}
\langle y \rangle & = & \frac{1}{M_{A}} \int d\bm{x} \langle
\overline{\psi} \gamma^{+} [i\partial^{+}+g_{v}V^{+}] \psi \rangle
\label{eq:avg-y-op}
\end{eqnarray}
Using the vector meson field equation in QHD-I
\[
\partial_{\mu}V^{\mu+}+m_{v}^{2}V^{+}=g_{v}\overline{\psi}
\gamma^{+}\psi,
\]
integrating by parts, and anti-symmetrizing one can re-express the
second term of Eq.~(\ref{eq:avg-y-op})
\begin{eqnarray}
\langle y \rangle & = & \frac{1}{M_{A}} \int d\bm{x} \langle
\overline{\psi}\gamma^{+} i\partial^{+} \psi +
m_{v}^{2}V^{+}V^{+}+V^{+\mu}V_{\mu}^{\:\:+} \rangle\nonumber\\* &
= & \frac{1}{M_{A}} \int d\bm{x} \langle  T^{++} -
\partial^{+} \phi \partial^{+}\phi \rangle\nonumber\\*
& = & \frac{1}{M_{A}}\left( P^{+}-P^{+}_{s} \right)\nonumber\\* &
= & 1 - \frac{P^{+}_{s}}{M_{A}} \simeq 1 \label{eq:HvH}
\end{eqnarray}
where $T^{++}$ is the canonical energy momentum tensor,
$P^{+}_{s}$ is the plus momentum of the scalar meson fields, and
$P^{+}$ is the total nuclear plus momentum. The result
Eq.~(\ref{eq:HvH}) constitutes an analog of the Hugenholtz-van
Hove theorem \cite{HvH} for finite systems; the equality becomes
exact in the nuclear matter limit, where the scalar meson
contribution vanishes, as shown in our previous work
\cite{Miller:2001tg}. This means that we may anticipate that the
binding effect will again be small. The `mixing' of the vector
operators and the scalar meson contribution will be elaborated on
in a more general context in Section \ref{sec:scalar}.

It is also worthwhile to explicitly evaluate the expression
Eq.~(\ref{eq:fn-general}) for $f_{N}(y)$ in the limit of infinite
nuclear matter. In this case, $V^{0}=V^{+}=V^{-}$ are constant and
$\bm{V}_{\bot}=0$, so we find
\begin{eqnarray}
\Lambda(z,\bm{x}_{\bot}) & = & \int_{z}^{\infty}dz'
V^{0}(z',\bm{x}_{\bot})\nonumber\\
& = & -V^{0}z\nonumber\\
& = & -V^{+}z,
\end{eqnarray}
so that Eq.~(\ref{eq:tranpsi}) becomes
\begin{eqnarray}
\langle \bm{k} |\psi''^{+}_{\alpha}\rangle & = & \int d\bm{x}
e^{-i\bm{k}_{\bot}\cdot\bm{x}_{\bot}} e^{-i(k^{+}-g_{v}V^{+})z}
\langle \bm{x} |\psi^{+}_{\alpha}\rangle\nonumber\\
& = & \langle \bm{k}_{\bot}, k^{+}-g_{v}V^{+}
|\psi^{+}_{\alpha}\rangle.
\end{eqnarray}
Therefore Eq.~(\ref{eq:fn-general}) becomes
\begin{eqnarray}
f_N(y) & = & 2\overline{M} \sum_{\alpha\in F} \int
d^{2}\bm{x}_{\bot} \left|\langle \bm{x}_{\bot},
\overline{M}y-g_{v}V^{+} |\psi^{+}_{\alpha}\rangle \right|^{2},
\end{eqnarray}
which is simply the expression~(\ref{eq:flf}) modified by a shift
in the argument of $g_{v}V^{+}/\overline{M}$. Thus  we find
Eq.~(\ref{eq:shift-nm}) is satisfied in the nuclear matter limit.
It is important to stress that all that is recovered here is the
shift in the argument and not any particular form of the plus
momentum distribution which arises from the specific model used.

\section{Nuclear Structure Functions}
\label{sec:numbers}

\begin{figure}
\centering
\includegraphics[scale=1.00]{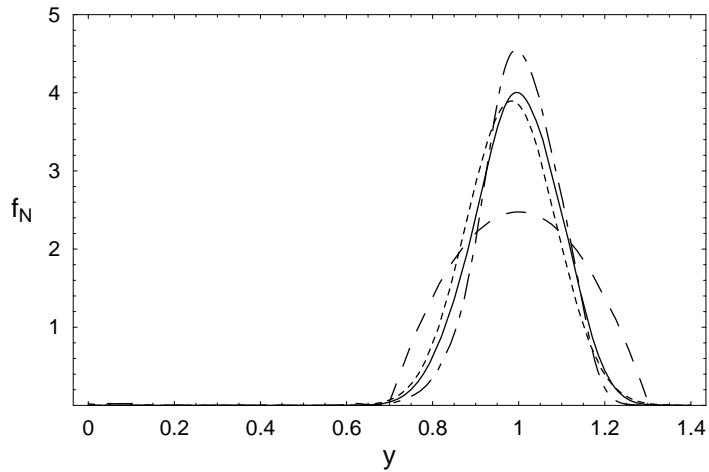}
\caption{Plus momentum distributions, $f_{N}(y)$, for $^{40}$Ca
(solid), $^{16}$O (short dashes), $^{208}$Pb (dot-dashes) and
nuclear matter (long dashes).} \label{fig:fn}
\end{figure}

\begin{figure}
\centering
\includegraphics[scale=1.00]{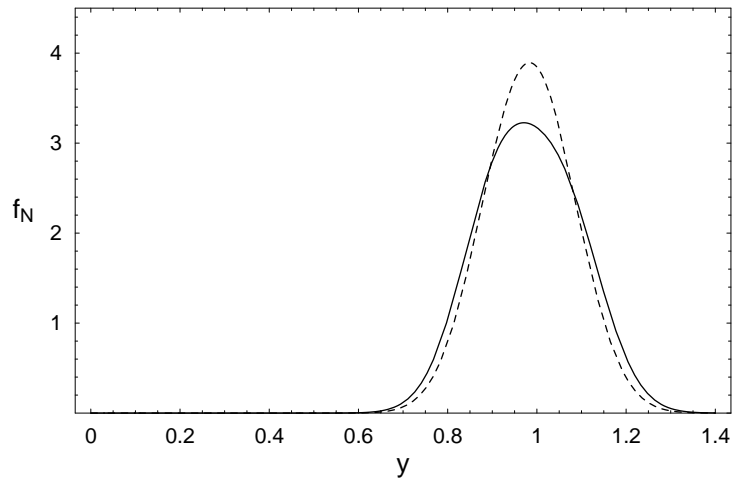}
\caption{Plus momentum distribution for $^{16}$O calculated with
harmonic oscillator (solid curve) and relativistic Hartree (dashed
curve) wave functions.} \label{fig:HOvsRH}
\end{figure}

We determine the wave functions appearing in
Eq.~(\ref{eq:fn-general}) numerically from a relativistic
self-consistent treatment following Horowitz and Serot
\cite{Horowitz:1981xw} using the same program \cite{cjht} which
includes electromagnetic effects. The plus momentum distribution
follows and is given in Fig.~\ref{fig:fn} for $^{16}$O, $^{40}$Ca,
$^{208}$Pb and in the nuclear matter limit (the $^{16}$O
calculation is also shown in Fig.~\ref{fig:HOvsRH}). One can see
that the peaks appear near $y=1$ as required by the Hugenholtz-van
Hove theorem Eq.~(\ref{eq:HvH}).

\begin{figure}
\centering
\includegraphics[scale=1.00]{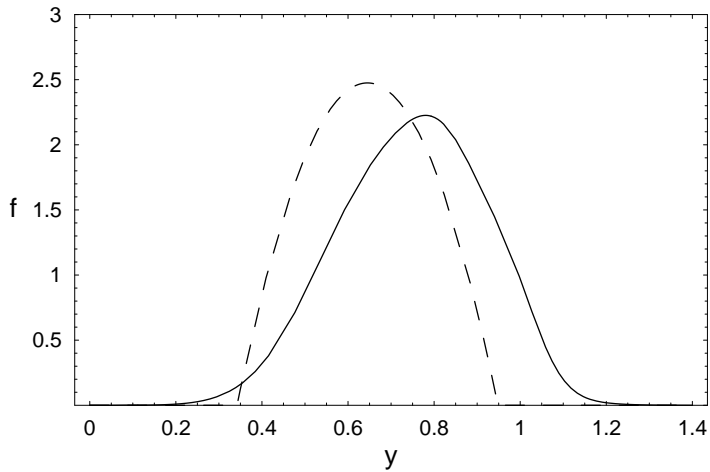}
\caption{$f(y)$ distributions for $^{16}$O and nuclear matter
after application of the Soper-Yan transformation. Note that the
peaks occur at $y<1$.} \label{fig:fblunden}
\end{figure}

It is worth noting that application of the Soper-Yan
transformation Eq.~(\ref{eq:soperyan}) to the $\psi''$
wavefunctions obtained from the equal time wavefunctions
reproduces the plus momentum distributions, including the correct
asymmetry, of the light front calculations in
Ref.~\cite{Blunden:1999gq}, which did not use the approximation
Eq.~(\ref{eq:prime-doubleprime}), as shown in
Fig.~\ref{fig:fblunden} for Oxygen and nuclear matter.
This demonstrates the excellence of the approximation relating
light front and equal time wavefunctions. One can see that the
effect in finite nuclei of the Soper-Yan transformation is to
shift and broaden the plus momentum distribution, while in nuclear
matter it is just a shift. If these distributions were to be used
in the nuclear structure function Eq.~(\ref{eq:convolution})
though, since $\langle y\rangle\simeq 0.8$ for Oxygen, the ratio
function Eq.~(\ref{eq:ratio}) would fall precipitously to nearly
zero at $x\simeq 0.6$ in stark contradiction with experiment.

We also evaluated the plus momentum distribution
Eq.~(\ref{eq:fn-general}) with the simple non-relativistic
harmonic oscillator shell model as an additional check on our
method. These (equal time) wavefunctions give us an explicit,
although approximate in the sense of Section \ref{sec:subtle},
closed form of the plus momentum distribution for $^{16}$O:
\begin{eqnarray}
f_N(y) & \simeq & \frac{\xi}{\sqrt{16\pi}}
[e^{-\xi^{2}(\eta_{1s}-y)^{2}}\nonumber \\* & & +
2\left(1+\xi^{2}(\eta_{1p}-y)^{2}\right)e^{-\xi^{2}(\eta_{1p}-y)^{2}}]
\label{eq:fnHO},
\end{eqnarray}
with $\eta_{nl} =M  + \omega(2(n-1)+l+3/2)-v_{0}$ where $v_{0}
\simeq 50 \text{ MeV}$, and $\xi \equiv \overline{M}b$ where
$b=(m\omega)^{-1/2} \simeq 1.6\text{ fm}$ is the oscillator length
which is fit to the root mean square radius of Oxygen $\langle
R^{2} \rangle^{1/2} \simeq 2.7 \text{ fm}$. The distribution
Eq.~(\ref{eq:fnHO}) narrows for larger $\xi$ which corresponds to
an increasing root mean square radius. This distribution is
plotted in Fig.~\ref{fig:HOvsRH} where one can see that it peaks
near $y=1$ like the relativistic Hartree calculation, but appears
to have a smaller value of $\langle y\rangle$.
It is worth noting that the Hartree calculations are in the
relativistic equal time framework and put into our relativistic
light front formalism, while the harmonic oscillator calculations
are non-relativistic and put into our relativistic formalism.

\begin{figure*}
\centering
\includegraphics[scale=1.0]{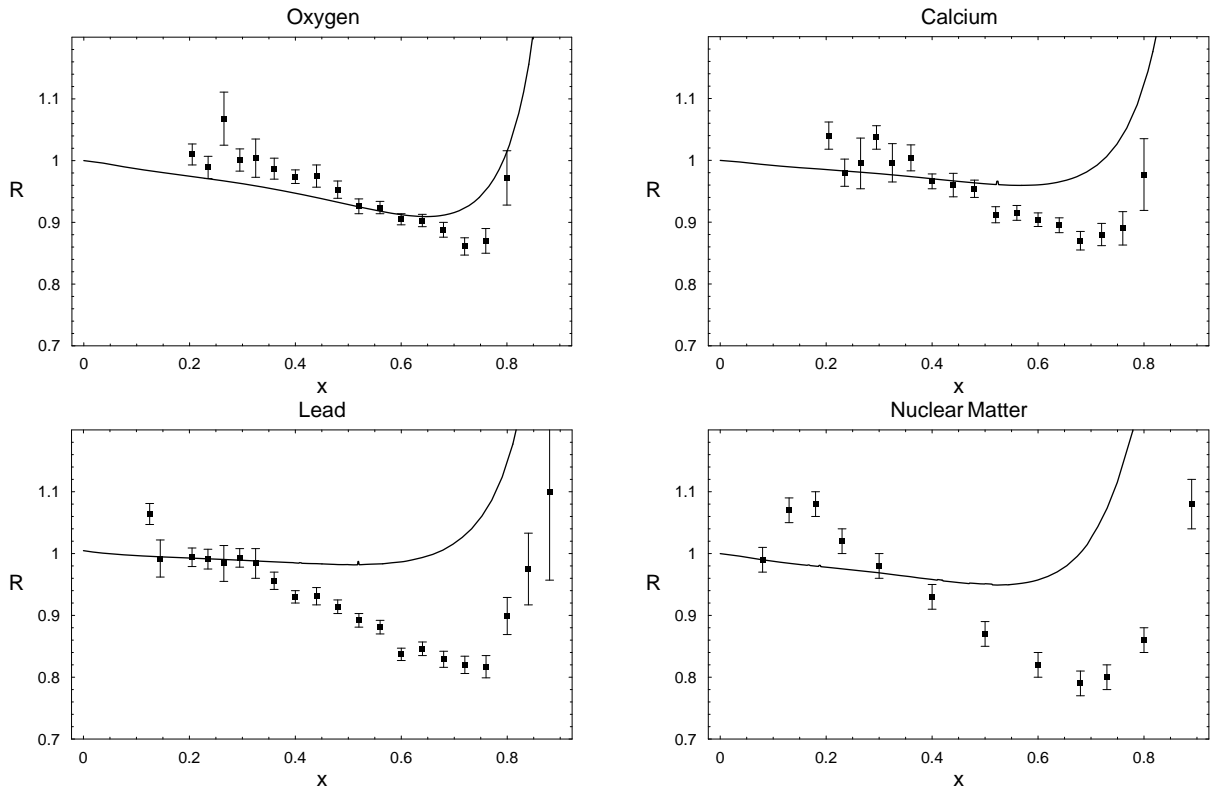}
\caption{Ratio functions for $^{16}$O, $^{40}$Ca and $^{208}$Pb
showing data for Carbon, Calcium and Gold, respectively, from
SLAC-E139 \cite{Gomez:1993ri}. The nuclear matter calculation
shows extrapolated data \cite{Sick:1992pw}.} \label{fig:ratio}
\end{figure*}

The structure function is given by the convolution
\begin{eqnarray}
\frac{F_{2A}(x_{A})}{A}=\int_{x_{A}}^{A} dy f_{N}(y)
F_{2N}(x_{A}/y), \label{eq:convolution}
\end{eqnarray}
with $x_{A}\equiv Q^{2}A/2P\cdot q = xM/\overline{M}$. The
assumption that nuclear effects do not modify the structure of the
nucleon is embodied in Eq.~(\ref{eq:convolution}) by the use of
the structure function of a free nucleon; we use the
parameterization \cite{deGroot:yb}
\begin{equation}
F_{2N}(x)=0.58\sqrt{x}(1-x)^{2.8}+0.33\sqrt{x}(1-x)^{3.8}+0.49(1-x)^{8}.
\end{equation}
The experiments measure the ratio function, defined as
\begin{equation}
R(x)=\frac{F_{2A}(x_{A})}{AF_{2N}(x)} \label{eq:ratio}.
\end{equation}
The results of our calculations are plotted for $^{16}$O,
$^{40}$Ca, $^{208}$Pb and in the nuclear matter limit in
Fig.~\ref{fig:ratio} showing data for Carbon, Calcium and Gold
from SLAC-E139 \cite{Gomez:1993ri} and an extrapolation
\cite{Sick:1992pw} for the nuclear matter calculation.
The most striking result is that these calculations fail to
reproduce the EMC effect; the curves consistently miss the minima
in the data, and the agreement gets worse with increasing $A$.
Another important result is that the ratio function does not fall
to zero as would be the case if the small effective mass ($\sim
0.56 M$ for nuclear matter in QHD-I) were the relevant parameter
describing the binding effect which would follow from using
Eq.~(\ref{eq:flf}) instead of Eq.~(\ref{eq:fn-general}). The
results also show a minimum near $x\simeq 0.6$ for Oxygen and
nuclear matter that is deeper than the Calcium and Lead
calculations. This is a curious feature that contradicts the trend
in experimental data, and is due to the effects of two parameters.

The first, and most important, is that of the location of the peak
of the plus momentum distribution given by Eq.~(\ref{eq:HvH}),
which gradually approaches $y=1$ as the nuclear matter limit is
reached. This is due to the fact that scalar mesons carry a small
amount of plus momentum \cite{Blunden:1999gq} that vanishes as
$A\rightarrow\infty$. The closer to $y=1$ the peak is in
Fig.~\ref{fig:fn}, the less pronounced the minimum in
Fig.~\ref{fig:ratio}, all else remaining constant. The second
effect is due to $\overline{M}$, which reaches a minimum at
$^{56}$Fe corresponding to a more pronounced minimum of the ratio
function than for $A<56$ or $A>56$, keeping the scalar meson
contribution constant.

\begin{figure}
\centering
\includegraphics[scale=1.00]{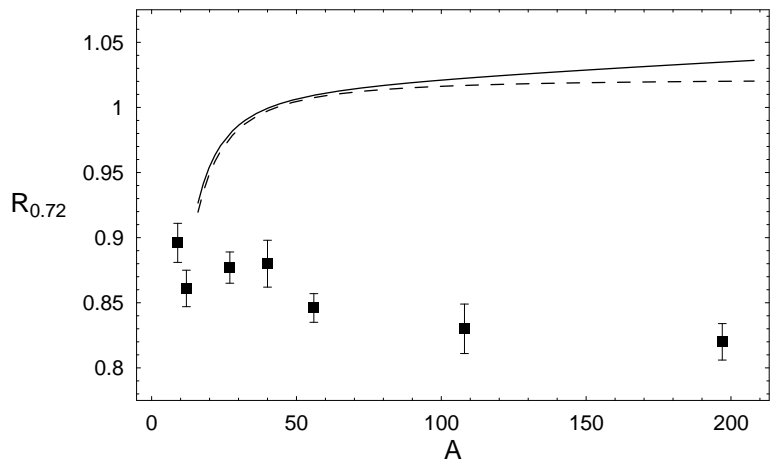}
\caption{$R(x=0.72)$ as a function of $A$ including scalar meson
and binding effects (solid line), and leaving binding energy per
nucleon constant at $-8.5\text{ MeV}$ (dashed line). The data are
from SLAC-E139 \cite{Gomez:1993ri}.} \label{fig:rminimum}
\end{figure}

Using a Gaussian parameterization of the plus momentum
distribution and the experimental binding energy per nucleon via
the semi-empirical mass formula, we have modeled the dependence of
the minimum of the ratio function, $R(x\simeq0.72)$, on the number
of nucleons in the nucleus in Fig.~\ref{fig:rminimum}.
The motivation for the use of Gaussian plus momentum distributions
is based on the expansion \cite{Frankfurt:nt}
\begin{eqnarray}
F_{2A}(x_{A}) & = & F_{2N}(x_{A})+\epsilon x_{A}
F_{2N}'(x_{A})\nonumber\\* & & + \gamma [2 x_{A}F_{2N}'(x_{A}) +
x_{A}^{2}F_{2N}''(x_{A})]\\* \text{where } \epsilon & \equiv &
1-\int dy y f_{N}(y)\\* \gamma & \equiv & \int dy (y-1)^{2}
f_{N}(y)
\end{eqnarray}
The Gaussian parameterization uses the peak location and width,
$\langle y\rangle$ and $(\langle y^{2}\rangle-\langle
y\rangle^{2})^{1/2}$ respectively, from the relativistic Hartree
calculations in Fig.~\ref{fig:fn}, and is normalized to unity.
This allows us to obtain a plus momentum distribution for any $A$
with minimal effort. We show the combined effect of scalar mesons
and binding energy per nucleon on the ratio function along with
the effect of scalar mesons alone using a constant binding energy
per nucleon of $-8.5\text{ MeV}$ independent of $A$. It can be
seen that a changing $\overline{M}$ with $A$ has the most effect
for nuclei much larger than Iron, but does not change the general
trend that the minimum of the ratio function becomes less
pronounced as $A$ increases due to the vanishing scalar meson
contribution and the peak of the plus momentum distribution
approaching unity. This dependence of the binding effect on $A$ is
quite different, both in magnitude and shape, than the trend in
experimental data summarized in Ref.~\cite{Sick:1992pw} which
satisfies $R(x\simeq 0.72)\sim A^{-1/3}$, so that the minimum
becomes more pronounced as $A$ increases. This fully demonstrates
the inadequacy of conventional nucleon-meson dynamics to explain
the EMC effect.

\section{Scalar Meson Contribution to Plus Momentum and More General Considerations}
\label{sec:scalar}

The average value of $y$, given by Eq.~(\ref{eq:HvH}), yields the
nucleon contribution to the plus momentum, and is less than one
which can be seen in Fig.~\ref{fig:fn}. We now address the
remaining plus momentum in finite nuclei. Previous results
\cite{Blunden:1999gq} show that a small fraction ($\delta y\sim
0.005$) of the plus momentum is carried by the scalar mesons which
vanishes as the nuclear matter limit is approached. This is due to
the fact that scalar mesons couple to gradients in the scalar
density (arising mainly from  the surface of finite nuclei) which
vanish as $A\rightarrow\infty$. The question is: why are scalar
mesons allowed to carry plus momentum and not vector mesons?

The simplest answer lies in the Dirac structure of
Eq.~(\ref{eq:fn}); the $\gamma^{+}$ in the trace picks out terms
in the full interacting Green's function with an odd number of
gamma matrices which includes all Lorentz vector interactions and
excludes Lorentz scalar interactions. The Dirac structure of
$f_{N}(y)$ is directly related to the Dirac structure of the
energy momentum tensor, so the answer also lies there and
illuminates a problem with conventional nucleon-meson dynamics.
The component of the energy momentum tensor relevant to the plus
momentum, from a chiral Lagrangian containing isoscalar vector
mesons, scalar mesons and pions, is given by
\cite{Miller:1997cr,Miller:1999ap}
\begin{eqnarray}
T^{++} & = & V^{+\mu}V_{\mu}^{\:\: +} + m_{v}^{2}V^{+}V^{+} +
\overline{\psi}\gamma^{+}i\partial^{+}\psi
 + \partial^{+}\phi\partial^{+}\phi \nonumber\\*
& & + \partial^{+}\bm{\pi} \cdot
\partial^{+}\bm{\pi} + \bm{\pi}\cdot\partial^{+}\bm{\pi}
\frac{\bm{\pi}\cdot\partial^{+}\bm{\pi}}{\pi^{2}}
(1-\frac{f^{2}}{\pi^{2}}\sin^{2}\frac{\pi}{f}).\label{eq:t++}
\end{eqnarray}
Since each of the terms in Eq.~(\ref{eq:t++}) involves one of the
fields, it is natural to associate each term with a particular
contribution to the plus momentum. This decomposition, though, is
not well defined; field equations relate various components. We
see the first three terms of Eq.~(\ref{eq:t++}) appear in $\langle
y \rangle$, which defines the nucleon contribution to the total
nuclear plus momentum, in the derivation of the Hugenholtz-van
Hove theorem Eq.~(\ref{eq:HvH}); we are not allowed to have the
vector mesons contribute a well defined fraction of plus momentum.
This means that one could trade certain mesonic degrees of freedom
for nucleons by replacing mesonic vertices with nucleon point
couplings, for example, in line with the general concept of
effective field theory. In our case the first three terms are
related by the vector meson field equation, but the fourth is left
out since the scalar mesons couple to the scalar density
$\overline{\psi}\psi$ which is not present in Eq.~(\ref{eq:t++}).
Therefore the scalar mesons (and pions) contribute a well defined
fraction of plus momentum. These explicit meson contributions
create an EMC binding effect, but the pionic contributions  are
also limited by nuclear Drell-Yan experiments \cite{Alde:1990im}
to carrying about 2\% of the plus momentum which is insufficient
to account for the entire EMC effect which corresponds to about
5\% of the plus momentum for Iron.

\section{Summary and Discussion}
\label{sec:summary}

The minimum in the EMC effect is known to have a monotonically
decreasing behavior with $A$, which has been studied in
Refs.~\cite{Gomez:1993ri,Sick:1992pw} among others. Our present
theory is defined by the use of the mean-field approximation,
along with the assumption that nuclear effects do not modify the
structure of the nucleon. This theory leads to results in severe
disagreement with experiment. Not only do we find that the depth
of the minimum is monotonically decreasing with $A$, but it has a
smaller magnitude than experiment. These results, which fail to
capture any of the important features of the experiments,
represent a failure of relativistic mean field theory.
Furthermore, the plus momentum distributions we compute give
$\langle y\rangle\simeq 1$ which indicates that nearly all of the
plus momentum is carried by the nucleons. In order to reproduce
the data, the nucleon plus momentum must be decreased by some
mechanism that becomes more important at larger $A$.
Nucleon-nucleon correlations cannot take plus momentum from
nucleons, and explicit mesonic components in the nuclear Fock
state wavefunction carrying plus momentum are limited
\cite{Bickerstaff:1984ax, Ericson:1984vt, Berger:1986dr} by
Drell-Yan experiments \cite{Alde:1990im}. Thus it appears that the
EMC effect may be due to something outside of conventional
nucleon-meson dynamics. For example, true modifications to nucleon
structure caused by nuclear interactions could be important, in
which case one would need to use models such as the
mini-delocalization model \cite{Frankfurt:nt}, quark-meson
coupling (QMC) model
\cite{Saito:1994ki,Benhar:1999up,Gross:1992pi} or the chiral quark
soliton model reviewed in \cite{Diakonov:2000pa} to include those
effects.

\begin{acknowledgments}
We would like to thank the USDOE for partial support of this work.
\end{acknowledgments}

\end{document}